# Realization of a wide steering end-fire facet optical phased array using silicon rich silicon nitride

HANI NEJADRIAHI,[1] PRABHAV GAUR,[1] KARL JOHNSON,[1] STEVE PAPPERT,[1] YESHAIAHU FAINMAN[1], PAUL YU[1]

*Department of Electrical & Computer Engineering, University of California, San Diego, 9500 Gilman Drive, La Jolla, CA 92093, USA*

*Corresponding author: hnejadri@eng.ucsd.edu*

**The design, fabrication, and characterization of a 16-element optical phased array (OPA) using a high index (n = 3.1) silicon rich silicon nitride (SRN) is demonstrated. We present one-dimensional beam steering with end-fire facet antennas over a wide steering range of >115° at a fixed wavelength of 1525 nm. A spot size of 0.11° has been measured at boresight, consistent with theory. We demonstrate SRN as a viable material choice for chip-scale OPA applications due to its high thermo-optic coefficient, high optical power handling capacity due to negligible two-photon absorption (TPA), wide transparency window, low propagation loss, and CMOS compatibility.**

___________________________________________

## 1. Introduction

Chip based silicon photonics optical phased arrays (OPA) are an attractive alternative to replace the complex free-space optical systems used in LiDAR applications [1]. Two material choices used in optical phased arrays are silicon and stoichiometric silicon nitride. However, each of these materials have its own downsides for offering an ideal optical phased array/LiDAR platform. Silicon is mostly implemented in the silicon on insulator (SOI) wafers and silicon nitride is also commonly found in many CMOS foundries as a wave-guiding material alternative. However, silicon can only guide above its band gap wavelength at around 1.1 μm while silicon nitride is transparent at both visible and the near infrared wavelength regions. Usually, silicon photonics OPA/ LiDAR systems work around the telecom wavelengths (O and C bands) for eye safety reasons, however depending on the application of interest, this can change [1-4].

Commercial LiDAR systems, for example, operate at 904-940 nm due to less solar noise at these wavelengths compared to O- and C-bands. Silicon nitride's transparency in the 800 to 1100 nm spectral range allows for leveraging many light sources, and implementations for many applications that could not be achieved using silicon material platform. On the other hand, silicon nitride's lack of an efficient phase tuning leads to greater power consumption and larger footprint phase shifters. The optical propagation loss values shown in Table 1 are good measures of scalability for these two material platforms. While silicon's thermo-optic coefficient (1.8 x 10$^{-4}$) is an order of magnitude higher and allows for a relatively more efficient tuning, it is important to note that waveguides for optical phased arrays often need to transmit at very high power densities for which the nonlinear effects are not negligible and must be taken into account. Silicon has a relatively large TPA coefficient, $\beta_{TPA}$. Generally, $\beta_{TPA}$ can be used to describe the propagation loss as $10 log_{10}(e) \frac{\beta_{TPA}}{A_{eff}}$, where the $A_{eff}$ is the effective mode size area. In case of a silicon waveguide with cross-sectional area of (220 x 500 nm$^2$) at 1550 nm, we can assume on average the loss is around 2 dB/cm per 1 Watt of optical power, therefore, TPA introduces non-negligible loss when transmitting above 100's of mW of optical power - which is a bottleneck for practical implementation of OPA and their full integration in LiDAR system. Hence, our SRN based OPA with a larger transparency window starting at 700 nm, negligible TPA coefficient compared to crystalline silicon, high thermo-optic coefficient and high refractive index for compact and more efficient devices can be an alternative platform for OPA and LiDAR systems. [5,6].

**Table 1. A comparison of typical designs and properties between Si and SiN waveguide with SiO2 cladding**

|  | Transparency μm | Thickness μm | Loss dB/cm | Bending μm | $\frac{\beta_{TO}}{K}$ | n2 cm$^2$/GW | $\beta_{TPA}$ cm/GW |
|---|---|---|---|---|---|---|---|
| **Si** | 1.1-3.7 | 0.2-0.5 | 1-3 | 5-50 | 1.8 x 10$^{-4}$ | 5 x 10$^{-5}$ | 0.5 |
| **SiN** | 0.5-3.7 | 0.2-2 | 0.2-2 | 20-200 | 2-4 x 10$^{-5}$ | 3-7 x 10$^{-6}$ | negligible |

In this paper, we demonstrate the implementation of a one-dimensional 16-element SRN phased array with 115° field of view and over 80 percent of power in the single-diffracted-beam. The spot size is 0.11° at boresight. We further discuss in detail the antenna design, phase shifter implementation, and the use of gradient descent algorithm for phase control. Simultaneous control of 16 thermo-optic phase shifters required integration of the optical phased array chip onto a printed circuit board (PCB) for electrical control.

___________________________________________

*A. Antenna design and simulation results*

Despite the scalability challenges with OPAs, a large number of antennas lead to small beam widths. Here, we investigate a 16-element optical phased array to achieve a relatively large steering angle range while minimizing the scalability challenges. The spacing between the antenna elements is governed by the spacing between the emitting elements in the array and we keep our element spacing at a period λ/2.

Even though at this spacing, high beam efficiency is achievable, however, evanescent coupling between the neighboring waveguide emitters must be avoided. To do so, for long propagation lengths (over mm scale), we can minimize their overlap in the phase space by creating a mismatch in their β coefficients. We did so by designing and fabricating waveguides of different widths. Their widths were chosen carefully and were kept in a similar range of values to ensure uniform illumination across the array of emitters. The waveguides are phase mismatched with both their nearest neighbor and their second nearest neighbor (separated by λ/2 and λ, respectively) by cycling through a set of three widths (300, 400, 500 nm in a sequence). Also, keeping the height of all of these waveguides the same and at 320 nm, ensures single-mode for TE-polarized light across the entire propagation length. [7].

Lumerical EME simulation of the waveguide array with sequentially varying widths for a 1 mm length propagation direction showed minimal coupling. Additionally, we experimentally demonstrate that even for several-mm propagation lengths, coupling between nearest, second nearest, and third-nearest neighbor waveguides is below -18 to -22 dB and power propagates only in the waveguide into which light was originally launched.

*B. End-fire facet design and simulation for far field projection*

In a uniform linear array with N elements, the far-field pattern depends on both the phase relationships between individual elements and their amplitude uniformity. Here we present the simulation results of Finite Difference Time Domain (FDTD) of the end-fire facet waveguide antennas shown in Fig. 1a.

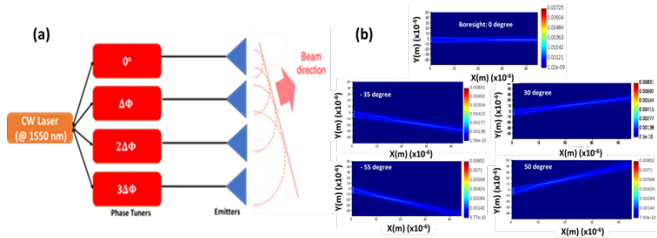

Figure 1. (a) The phased array schematic for the FDTD simulation where the far field beam profile is analyzed based on the different phase patterns. (b) The E-field intensity profile of the steered beam at boresight, -30°, 30°, -55°, and 50°

From Fig.1, we observe that by varying the phase of the individual waveguiding emitters, we achieve the far field intensity profile of the 16-element array for a few selected steering angles, Fig. 1.b An observation can be made that for steering angles closer to the physical limit of the array (~ +/-60°), the beam width increases (albeit hard to quantify in this plot).

To better understand the far field characteristics of the emitters array two cases were studied. In the first case, all the waveguides are set to 400 nm width and 320 nm height placed at a λ/2 spacing (775 nm). The simulation was set up so as to ensure far field propagation. Fig. 2.a shows the individually normalized intensity profiles of specific target angles of the full array overlaid with the element factors of three different waveguide widths of interest, namely 300, 400, and 500 nm. It can be observed from Fig. 2.a that the far field intensity profile of these three individual waveguides is very similar, and the aperture of the array results in a uniform Gaussian envelope profile. The same study was repeated to ensure that the performance remains unchanged in the case of nonuniform waveguide antenna arrays. From Fig. 2.b it can be concluded that despite the negligible difference between the waveguide's widths, the overall aperture of the array remains unchanged. While OPA configurations can differ in terms of platforms, architecture, and components, however the key metrics for the performance evaluation or the beam forming quality, i.e., far field aperture (field of view) and the aliasing free steering range, solely depend on the geometrical properties of the emitters.

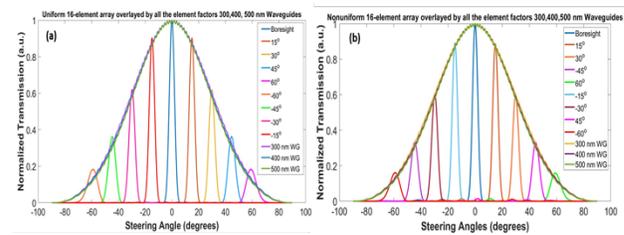

Figure 2. a) The far field pattern of individual waveguide elements with uniform antenna width (for the case where all waveguides are all either 300, 400, or 500 nm in width) overlaid with the antenna array at different steering angles b) Similar to (a) the far field pattern of individual waveguide elements with nonuniform antenna width

Ideally, to divide the light on a chip we would start with a series of cascaded 1xN splitters. Here we use 1x2 Multi Mode Interferometer (MMI) splitter and cascade them with 1x8 MMI splitters to distribute the light into 16 segments. The splitter's length and width are designed in such a way that when the higher order modes are excited the input electric field is self-imaged to the outputs. Tapered widths are not only useful for compactness, but also allow for an increase in the light intensity near the dielectric corners with abrupt discontinuities and thus reduce reflections. The optimized design ideally needs to have low insertion loss and low power imbalance between the output waveguides, so the figure of merit defined for the Particle Swarm optimization is as follows: $FOM = \frac{\sum_{i=1}^{8} c_i}{(0.1 + \max(c_i) - \min(c_i))}$ where $c_i$ is the power transmission from the input waveguide to the output waveguide #i. Hence with that FOM, we use coupler length, coupler width, taper length, taper width, and the gap as parameters for the optimization. In the case of the 1x2 MMI coupler, the insertion loss is 0.021 dB. In the case of the 1x8 MMI, the transmission spectra of the simulated MMI coupler show a power imbalance over the 0.1 μm wavelength range. The wavelength of operation is best to be chosen at 1.525 μm since the output powers are identical - this can be seen from the transmission spectra shown in Fig. 3. The insertion loss of 0.55 dB is calculated. Using an MMI coupler introduces the challenges of proper phase correction and operation at a wavelength where the optical outputs are equal. However, implementing them helps with minimizing propagation loss in comparison to the more traditional designs such as a y-branch coupler, especially for optical systems where scaling is key (i.e., OPAs). [7,10]. The Optical Microscope (OM) image of the fabricated coupler is shown in Fig. 3.d, and 3.e shows the IR camera image of the 8 outputs from the coupler during a wavelength sweep. Here the wavelength was at 1525 nm and shows consistency with the simulated results.

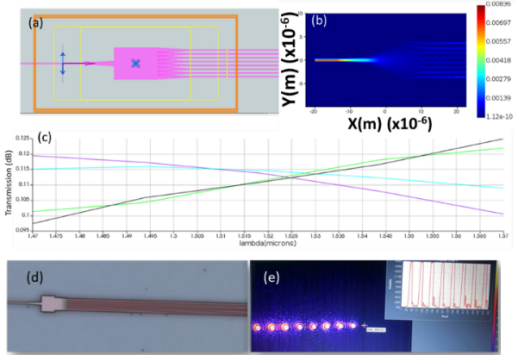

Figure 3. (a) The zoomed-in schematic of the 1 x 8 MMI coupler (b) The electric field intensity profile of the coupler (c) the transmission spectra of the simulated (d) The optical microscope image of the 1 x 8 MMI coupler (e) The optical modes of all 8 outputs measured and imaged with an Infrared camera

Further after splitting, the antennas are connected to a 1.2 mm long phase shifter as shown previously here [8-11]. These localized heaters used for the phase shifter are designed in such a way as to have the lowest minimum thermal crosstalk. Gold contact pads of size 200 x 200 μm$^2$ located 5 mm from the edge of the photonic chip for further wire-bonding purposes were also designed (see Fig. 4).

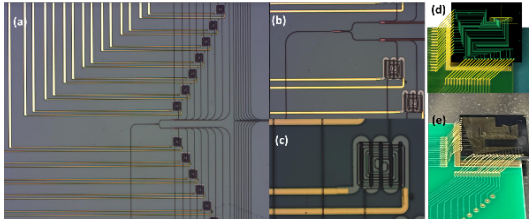

Figure 4. (a) The OM image of OPA chip (b) MMI splitters and phase shifters (c) zoomed-in OM image of the phase shifter with the contact pads attached (d & e) Schematic and the OM image of the OPA chip wire-bonded to the PCB- pre and post-adhesion process

After the phase shifters, the waveguides are routed in a 90° bend to form the final end-fire facet waveguides. The layout of the overall OPA design and the OM images of the actual fabricated chip with the zoomed-in MMI, phase shifter section, and the closely spaced waveguides are shown in Fig. 4.a-c. Bending radii ranging between of 15 to 85 μm were used to minimize coupling as the waveguides are brought close to each other and to avoid a significant change in their effective index and the loss of modal shape. The 16-waveguides now brought together at λ/2 spacing propagate for a distance of ~ 2 mm and then terminate close to the edge of the chip where they are diced and polished using a focused ion beam (FIB). After the facets are exposed, the waveguides act as emitters following a Gaussian pattern.

While we did not do any amplitude control of the waveguide antennas, it is important to note that we could replace the phase shifters with MMI switches shown [3] to achieve steering through amplitude modulation of the individual waveguide antennas. Further modifications of this chip could potentially involve a combination of both phase shifter and MMI switch to create a more uniform far field emission and or to improve the side lobe suppression. [8].

## 3. Phase control and active steering

To actively control the phase associated with each emitting element in the array, the photonic chip is first mounted and wire-bonded to a PCB. Figs. 4.d-e shows the pre- and post-epoxy covering of the wire-bonds. We use the gradient descent algorithm (GDA) to find the maximum intensity that could be achieved for a set of phase/voltage solutions. In this algorithm, we use an iterative optimization mechanism for finding the local maxima of the intensity function. Using the optimization algorithm and the feedback from the IR camera, a set of voltage values corresponding to different steering angles are obtained. This process is repeated until all the proper phases and their corresponding voltage values associated with each emitter required for the range of steering angles of interest are achieved [11,12].

After the accurate phase variations/corrections were implemented using the GDA, we were able to achieve steering at a fixed wavelength of 1525 nm. This wavelength was chosen due to the experimentally achieved equal power values of the MMI coupler outputs with minimal phase corrections. The far field image at a few different target angles is captured. Figs. 5.a-c are based on a smoothened and averaged out grey scale image of the longitudinal beams. The corresponding phase values for each case are also shown. The phases of the two opposite angles are symmetric in nature, albeit in an experiment can be altered according to the modifications required after fabrication imperfections [13,14].

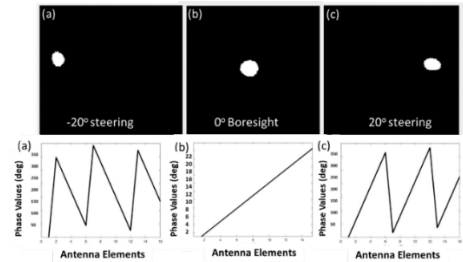

Figure 5. The far field images at 1525 nm after phase correction for -20°, 0°, 20° steering angles along with their corresponding phases

We experimentally achieve larger than 115° field of view. Fig. 6 is the far field beam patterns for the beam steered ±57.5° on-axis. Our array emits a high efficiency beam with 11.5 dB peak to sidelobe ratio which is very close to the theoretical limit of 13 dB (sinc$^2$ function coming from the far field pattern of a rectangular aperture). Further optimization can be done to suppress the sidelobes using different mechanisms i.e., nonuniform array spacing, etc.

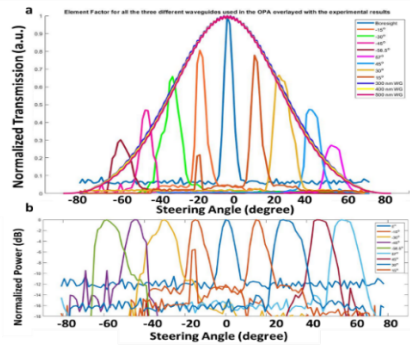

Figure 6. (a) Measured far field optical power as a function of the steering angle over a 120° field of view, each array factor has been normalized to the power value at boresight (0°)- the element factor of the three contributing waveguides have been overlayed on top showing the far field aperture (b) logarithmic scale plot of the beam steering normal to the array output showing peak to sidelobe ratios of 11.5 dB- the individual beams are normalized to themselves

Fig. 6.a shows the measured far field normalized power versus the steering angle over a large 120° field of view for different target beam angles. Here, the power is normalized to the power at boresight (0° peak) and the amplitudes scale accordingly. Note that we have also plotted the element factor from the emission of a single antenna waveguide (albeit for all the three different widths used in the antenna array i.e., 300, 400, and 500 nm) which closely resembles that predicted from the model shown in Fig 2. Fig. 6.b shows the logscale plot of the same, however in this case the power values are not normalized to the power value at boresight. The spot size of the measured OPA is ~ 0.11° at 0° beam. Although as expected the full width half maximum of the beam widens as it goes away from the boresight and is doubled at ±57.5°. The residual sidelobes apparent in the plots above are due to the limitation in the accurate phase manipulation of the integrated phase shifters and the phase uncertainty present in each corresponding waveguiding antenna [15,16].

## 5. Conclusion

In summary, we demonstrated that optical beam steering can be achieved in a new platform of SRN where we overcome the limitations of both silicon and stoichiometric silicon nitride material platforms. We used the design and fabrication of the PCB wire bonded to the photonic chip along with the gradient descent algorithm to accurately control the phase of the array. We show approximately 120° field of view in 1D and a spot size of 0.11° at boresight. Our component designs show high beam quality with negligible crosstalk between each phase shifters allowing for scaling to large number of antenna elements. We believe SRN can prove to be an excellent materials candidate for chip-scale OPA applications due to its wide transparency window, high thermo-optic effect, high power handling capability, and its low waveguide propagation loss.

**Funding.** National Science Foundation (NSF ECCS-2023730, ECCS-1542148); Office of Science (DE-SC0019273); Office of Naval Research (N00014-18-1-2027).

**Acknowledgments.** We thank UCSD's Nano3 cleanroom staff and especially Dr. Maribel Montero, for their assistance in the preparation of the samples.). The SRN synthesis is funded as part of the Quantum Materials for Energy Efficient Neuromorphic Computing, an Energy Frontier Research Center funded by the U.S. Department of Energy, Office of Science, Basic Energy Sciences (DE-SC0019273).

**Disclosures.** The authors declare no conflicts of interest.

## References

1. S.T. Ilie, T.D. Bucio, T. Rutirawut, L. Mastronardi, I. Skandalos, H. Chong, F.Y. and Gardes, 2021, March. Silicon-rich silicon nitride CMOS platform for integrated optical phased arrays. In Smart Photonic and Optoelectronic Integrated Circuits XXIII (Vol. 11690, p. 1169005). International Society for Optics and Photonics.
2. Q. Wang, S. Wang, L. Jia, Y. Cai, W. Yue, and M. Yu, 2021. Silicon nitride assisted 1× 64 optical phased array based on a SOI platform. Optics Express, 29(7), pp.10509-10517.
3. H.Nejadriahi, A.Friedman, R.Sharma, S.Pappert, Y.Fainman, and P.Yu, "Thermo-optic properties of silicon-rich silicon nitride for on-chip applications," Opt. Express 28, 24951-24960 (2020).
4. A. Marinins, S.P. Dwivedi, J.Ø. Kjellman, S. Kerman, T. David, B. Figeys, R. Jansen, D.S. Tezcan, X. Rottenberg, P. and Soussan, 2020. Silicon photonics co-integrated with silicon nitride for optical phased arrays. Japanese Journal of Applied Physics, 59(SG), p.SGGE02.
5. X. Sun, L. Zhang, Q. Zhang, and W. Zhang, 2019. Si photonics for practical LiDAR solutions. Applied Sciences, 9(20), p.4225.
6. Q. Wang, S. Wang, L. Jia, Y. Cai, W. Yue, and M. Yu, 2021. Silicon nitride assisted 1× 64 optical phased array based on a SOI platform. Optics Express, 29(7), pp.10509-10517.
7. Y. Oda, M. Kiso, S. Kurosaka, A. Okada, K. Kitajima, S. Hashimoto, G. Milad, and D. Gudeczauskas, 2008, November. Study of suitable palladium and gold thickness in ENEPIG deposits for lead free soldering and gold wire bonding. In ECTC.
8. H. Nejadriahi, S. Pappert, Y. Fainman, and P. Yu, "Efficient and compact thermo-optic phase shifter in silicon-rich silicon nitride," Opt. Lett. 46, 4646-4649 (2021).
9. A. Maese-Novo, R. Halir, S. Romero-García, D. Pérez-Galacho, L. Zavargo-Peche, A. Ortega-Moñux, I. Molina-Fernández, J.G. Wangüemert-Pérez, and P. Cheben, 2013. Wavelength independent multimode interference coupler. Optics express, 21(6), pp.7033-7040.
10. M. Bachmann, P.A. Besse, and H. Melchior, 1994. General self-imaging properties in N× N multimode interference couplers including phase relations. Applied optics, 33(18), pp.3905-3911.
11. F. Bagci,and B. Akaoglu, 2013. A 1x4 power-splitter based on photonic crystal Y-splitter and directional couplers. Opt. Pura Opt. Pura Apl, 46(3), pp.265-273.
12. P. Baldi, 1995. Gradient descent learning algorithm overview: A general dynamical systems perspective. IEEE Transactions on neural networks, 6(1), pp.182-195.
13. H.A. Clevenson, S.J. Spector, L. Benney, M.G. Moebius, J. Brown, A. Hare, A. Huang, J. Mlynarczyk, C.V. Poulton, E. Hosseini, and M.R. Watts, 2020. Incoherent light imaging using an optical phased array. Applied Physics Letters, 116(3), p.031105.
14. J. He, T. Dong, and Y. Xu, 2020. Review of photonic integrated optical phased arrays for space optical communication. IEEE Access, 8, pp.188284-188298.
15. A. Yaacobi, J. Sun, M. Moresco, G. Leake, D. Coolbaugh, and M.R. Watts, 2014. Integrated phased array for wide-angle beam steering. Optics letters, 39(15), pp.4575-4578.
16. H. Wang, Z. Chen, C. Sun, S. Deng, X. Tang, L. Zhang, R. Jiang, W. Shi, Z. Chen, Z. Li, and A. Zhang, 2021. Broadband silicon nitride nanophotonic phased arrays for wide-angle beam steering. Optics Letters, 46(2), pp.286-289